\documentclass[twocolumn,english,aps,prb,twocolum,superscriptaddress,bibnotes,amsmath,amssymb,floatfix]{revtex4-2}
\usepackage[colorlinks=true,citecolor=blue,linkcolor=magenta]{hyperref}

\usepackage{graphicx}
\usepackage{epsfig}
\usepackage{color}
\usepackage{amssymb}
\usepackage{amsmath}
\usepackage{array}
\usepackage[loose]{units}
\usepackage{braket}
\usepackage{soul}
\usepackage{hyperref}
\usepackage{cleveref}
\usepackage[caption=false]{subfig}
\usepackage[english]{babel}
\usepackage{letltxmacro}
\usepackage{siunitx}
\usepackage{ dsfont }
\LetLtxMacro{\ORIGselectlanguage}{\selectlanguage}
\makeatletter
\DeclareRobustCommand{\selectlanguage}[1]{%
    \@ifundefined{alias@\string#1}
      {\ORIGselectlanguage{#1}}
      {\begingroup\edef\x{\endgroup
         \noexpand\ORIGselectlanguage{\@nameuse{alias@#1}}}\x}%
}

\newcommand{\GOE}{Georg-August-Universit\"{a}t G\"{o}ttingen, D-37077 G\"{o}ttingen, Germany}
\newcommand{\EPFL}{Swiss Federal Institute of Technology Lausanne (EPFL), CH-1015 Lausanne, Switzerland}
\newcommand{\MPIBPC}{Max Planck Institute for Biophysical Chemistry, D-37077 Göttingen, Germany}

\begin{document}

\title{Integrated photonics enables continuous-beam electron phase modulation}

\author{Jan-Wilke Henke}
\thanks{These authors contributed equally.}
\affiliation{\GOE}
\affiliation{\MPIBPC}
\author{Arslan Sajid Raja}
\thanks{These authors contributed equally.}
\affiliation{\EPFL}
\author{Armin Feist}
\affiliation{\GOE}
\affiliation{\MPIBPC}
\author{Guanhao Huang}
\affiliation{\EPFL}
\author{Germaine Arend}
\affiliation{\GOE}
\affiliation{\MPIBPC}
\author{Yujia Yang}
\affiliation{\EPFL}
\author{F.~Jasmin Kappert}
\affiliation{\GOE}
\affiliation{\MPIBPC}
\author{Rui Ning Wang}
\affiliation{\EPFL}
\author{Marcel Möller}
\affiliation{\GOE}
\affiliation{\MPIBPC}
\author{Jiahe Pan}
\affiliation{\EPFL}
\author{Junqiu Liu}
\affiliation{\EPFL}
\author{Ofer Kfir}
\affiliation{\GOE}
\affiliation{\MPIBPC}
\author{Claus Ropers}
\email{cropers@gwdg.de}
\affiliation{\GOE}
\affiliation{\MPIBPC}
\author{Tobias J. Kippenberg}
\email{tobias.kippenberg@epfl.ch}
\affiliation{\EPFL}
\maketitle
\date{\today}


\textbf{The ability to tailor laser light on a chip using integrated photonics has allowed for extensive control over fundamental light-matter interactions in manifold quantum systems including atoms~\cite{Thompson2013, Chang2018, Corzo2019}, trapped ions~\cite{Mehta2016, Mehta2020, Niffenegger2020}, quantum dots~\cite{Sollner2015, Lodahl2015, Liu2018_high, Norman2018}, and defect centers~\cite{Sipahigil2016}. Beams of free electrons, enabling high-resolution microscopy for decades, are increasingly becoming the subject of laser-based quantum manipulation and characterization \cite{Barwick2009, Yurtsever2012, Piazza2015, GarciadeAbajo2010, Vanacore2018, Priebe2017, Feist2020}. Using free-space optical excitation and intense short laser pulses, this has led to the observation of free-electron quantum walks \cite{Feist2015, Echternkamp2016}, the generation of attosecond electron pulses \cite{Priebe2017, Morimoto2018, Kozak2018a}, and holographic imaging of electromagnetic fields \cite{Madan2019}. Enhancing the interaction with electron beams through chip-based photonics \cite{England2014, Sapra2020} promises unique applications in nanoscale quantum control and sensing, but has yet to enter the realm of electron microscopy.
Here, we merge integrated photonics with electron microscopy, demonstrating coherent phase modulation of a continuous electron beam using a silicon nitride microresonator driven by a continuous-wave laser. The high-Q factor ($\mathbf{Q_0 \sim10^6}$) cavity enhancement and a waveguide designed for phase matching lead to efficient electron-light scattering at unprecedentedly low, few-microwatt optical powers. Specifically, we fully deplete the initial electron state (zero-loss peak) at a cavity-coupled power of $\mathbf{6~\text{\boldmath$\mu$} W}$ and create $\mathbf{>}$~500 photon sidebands on the electron spectrum for only 38~mW in the bus waveguide. Moreover, we demonstrate $\mathbf{\text{\boldmath$\mu$} eV}$ electron energy gain spectroscopy (EEGS). Providing simultaneous optical and electronic spectroscopy of the resonant cavity, the fiber-coupled photonic structures feature single-mode electron-light interaction with full control over the input and output channels. This approach establishes a versatile framework for exploring free-electron quantum optics~\cite{Kfir2019, DiGiulio2019, Pan2019, BenHayun2021}, with future developments in strong coupling, local quantum probing, and electron-photon entanglement. More broadly, our results highlight the ability of integrated photonics to efficiently interface free electrons and light. }

\begin{figure*}[!ht]
\center
\includegraphics{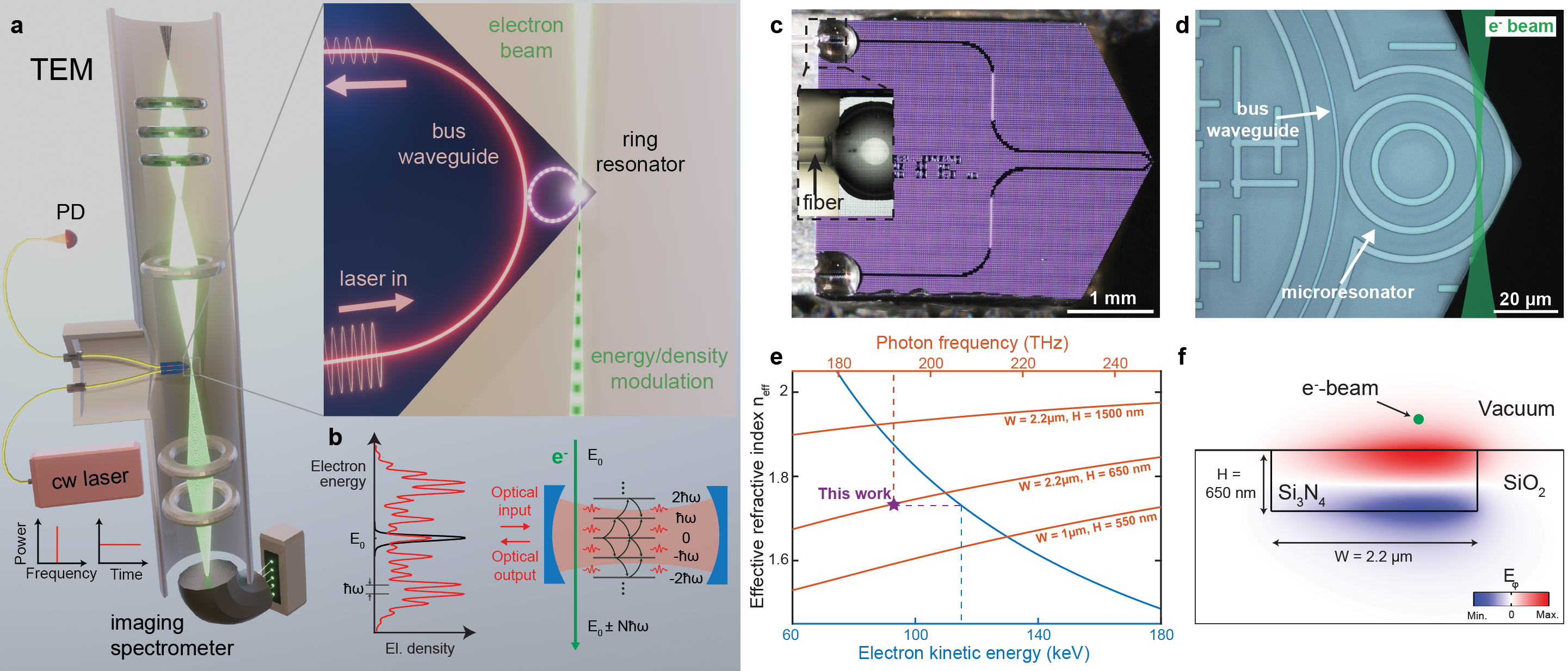}
\caption{\textbf{Principle of continuous-wave (CW) photonic-chip-based optical phase modulation of free-electron beams.} \textbf{a)} Rendering of the experimental setup including the electron source and a fiber coupled $\mathrm{Si_3N_4}$ photonic chip based microresonator. \textbf{b)} Left: electron spectra before (black) and after (red) interacting with the laser driven cavity mode. The electron-light interaction broadens the spectrum and generates discrete sidebands at integral multiples of the photon energy. Right: a cavity quantum electro-dynamical depiction of the electron-light interaction. The cavity photons induce transitions between the free-electron energy states. \textbf{c)} Photograph of the fiber-coupled $\mathrm{Si_3N_4}$ photonic chip mounted on a customized TEM holder. The triangular shaped chip edge minimizes undesired electron-substrate interactions. The inset shows an optical fiber glued to the input waveguide with a UV-curable epoxy. \textbf{d)} Magnified optical microscope image of the photonic chip showing the bus waveguide and the microresonator. The green path indicates the trajectory of the electron beam (not to scale) traversing the microresonator (parallel to the chip surface). \textbf{e)} Energy dependent effective index of the fundamental mode of the photonic microresonator and free electron. The integrated on-chip platform allows achieving phase matching at different electron kinetic energies (90-145 keV) by changing the width and height of the $\mathrm{Si_3N_4}$ waveguide. Phase matching is achieved between the optical mode at  $\sim$ 193.5 THz (corresponding to a wavelength of $\sim$ 1550 nm) and the free electrons at 115 keV by using a $\mathrm{Si_3N_4}$ waveguide having a dimension of $\sim$\SI{2.2}{ \micro m} $\times$ 650~nm. \textbf{f)} Finite element simulation of the $\mathrm{E_{\varphi}}$ distribution of the fundamental quasi-TM mode of microresonator. The green dot indicates the interaction location of the electron while the electron beam is propagating into the page.}\label{fig_setup}
\end{figure*}

\section*{Introduction}
The rapid advancement of electron microscopy epitomizes our growing ability to characterize the structure and function of nanoscale materials and devices. Based on disruptive improvements in spatial resolution and enabled by a variety of contrast mechanisms, electron beams serve as nearly universal probes in condensed matter research \cite{VanAert2011, Scott2012} and structural biology \cite{Cheng2018, Yip2020}. Beyond stationary imaging, the advent of in-situ and time-resolved electron microscopy  allow for the observation of transient phenomena and non-equilibrium dynamics \cite{Zewail2010}. Achieving temporal resolutions in the femtosecond regime \cite{Piazza2013, Bucker2016, Feist2017, Houdellier2018, Zhu2020}, ultrafast transmission electron microscopy (UTEM) has enabled stroboscopic real-space movies of numerous processes, including strain wave dynamics \cite{Yurtsever2009, Cremons2016, Feist2018}, ultrafast demagnetization \cite{Schliep2017, RubianodaSilva2018, Cao2021}, and the evolution of structural phase transformations \cite{vanderVeen2013, Danz2021}.
Moreover, in the form of photon-induced near-field electron microscopy (PINEM), UTEM permits quantitative, high-resolution imaging of nano-optical fields \cite{Barwick2009, Yurtsever2012, Piazza2015, Wang2020, Liebtrau2021}. The underlying process involves energy transfer between localized optical excitations and free electrons, modifying the state of the electron by generating discrete photon sidebands in a process termed stimulated inelastic electron-light scattering (IELS). The corresponding modulation of the electronic wavefunction at optical frequencies is quantum-coherent in nature \cite{Feist2015,GarciadeAbajo2010}, and thus can be used for the coherent control of electron quantum states in space \cite{Vanacore2018,Vanacore2019,Feist2020,GarciadeAbajo2021a} and time \cite{Priebe2017, Kozak2018a, Morimoto2018}. Recent theoretical work suggest that such modulated electron beams enable an electron-mediated transfer of optical coherence, predicting the generation of coherent cathodoluminescence, the resonant excitation of two-level systems, and superradiance from sequential electrons \cite{Gover2019, Zhao2020, DiGiulio2021, Kfir2021}. These developments imply a future close integration of electron microscopy with coherent optical spectroscopy, with opportunities in local quantum control and enhanced sensing.
A key obstacle in harnessing coherent electron-light interactions for scientific and technological applications is their present limitation to the ultrafast regime. With very few exceptions in IELS \cite{Das2019, Liu2019, Ryabov2020} and laser phase plates \cite{Schwartz2019}, accessing this regime has required timed electrons and high-intensity pulsed lasers to reach sufficient electron-photon coupling strength. Despite the use of phase matching and resonant amplification \cite{Peralta2013, Breuer2013, Kozak2017_acceleration, Dahan2020, Kfir2020}, achieving the continuous wave (CW) regime of electron phase modulation has remained out of reach of regular electron microscopes. 

Here, we overcome this challenge and demonstrate highly efficient electron-photon interactions in the continuous-wave regime, using an electron microscope and photonic integrated circuits based on $\mathrm{Si_3N_4}$. Our setup (Fig.~\ref{fig_setup}) allows for an electron beam to interact with the co-propagating evanescent field of a microresonator waveguide in the object plane of a transmission electron microscope (TEM).

\section*{Theory of cavity free-electron-photon interaction}

We first briefly review the quantum mechanical description of the electron-photon interaction that takes place \cite{Asenjo-Garcia2013}, adopting notions from cavity quantum electrodynamics (cQED) \cite{Kimble1998}, and applying it to our experiments in which electrons interact with evanescent field at a photonic chip (schematic in Fig.~\ref{fig_setup}).
The interaction between the electron beam and the ring cavity photons can be described by the Hamiltonian ~\cite{Asenjo-Garcia2013,Kfir2019,DiGiulio2019} (see derivation in SI)
\begin{equation}
H_1=\frac{e}{2m}(\hat p\cdot\hat A + \hat A \cdot\hat p)=\hbar g_0\hat a\hat b^\dag+\hbar g_0^*\hat a^\dag \hat b. 
\end{equation} 
While the length gauge is usually chosen for localized quantum systems in cQED, the above velocity gauge naturally describes free electrons at finite momentum. The interaction between the optical cavity (described by the annihilation operator $\hat a$) and a free electron (described by the electron-energy ladder operator $\hat b$, see SI) is described by the vacuum coupling rate
\begin{equation}
    g_0 =\eta \sqrt{\frac{e^2v_e^2}{2\epsilon\hbar\omega V}}.
\end{equation} 
Here $e$ is the electron charge, $v_e$ the electron group velocity, $\epsilon$ the optical permittivity, $\omega$ the optical frequency, and $V$ the effective optical mode volume. The phase matching condition is manifested in the coefficient $\eta = \int dze^{-i\Delta k\cdot z}u(z)/L$ defined by the optical mode profile function $u(z)$ along the electron trajectory and the electron wavenumber change $\Delta k=\omega / v_e$ upon photon absorption and emission~\cite{GarciadeAbajo2010,Park2010}. For an empty cavity, the interaction leads to electron-driven photon generation (cathodoluminescence) \cite{Bendana2011, Muller2021}. When driving the microresonator with a laser, the intracavity state is described by a coherent state $\ket{\alpha}$ such that the scattering matrix becomes \cite{Feist2015}
\begin{equation}
    S= \exp\left(g^* \hat b -g \hat b^\dag \right)~,
\end{equation}
introducing a dimensionless coupling constant $g = ig_0\tau\alpha$, where $|\alpha|^2=n_c$ is the mean intracavity photon number, and $\tau=L/v_e$ denotes the interaction time set by the transit time of the electron over the interaction region $L$. As a result of the interaction, prominent photon-sidebands in the electron kinetic energy spectrum are generated, separated from the initial energy $E_0$ by integer multiples of the photon energy $N\hbar\omega$ ($N\in\mathbb{Z}$) (Fig.~\ref{fig_setup}b). For a quasi-monochromatic electron, the photon sideband occupation probabilities $P_N$ can be expressed in the form of the $N$th-order Bessel function of the first kind, that is $P_N = J_N(2|g|)^2$.
Correspondingly, in the spatial representation of the electron state, the interaction with the field imprints a sinusoidal phase modulation onto the wavefunction. One can therefore view the interaction as phase modulation of electrons operating at optical frequencies. The coupling constant $g$ in principle is a complex number, for simplicity, however, in the following parts of the manuscript we use $g$ in place of $|g|$.

\begin{figure*}[!ht]
\centering
\includegraphics{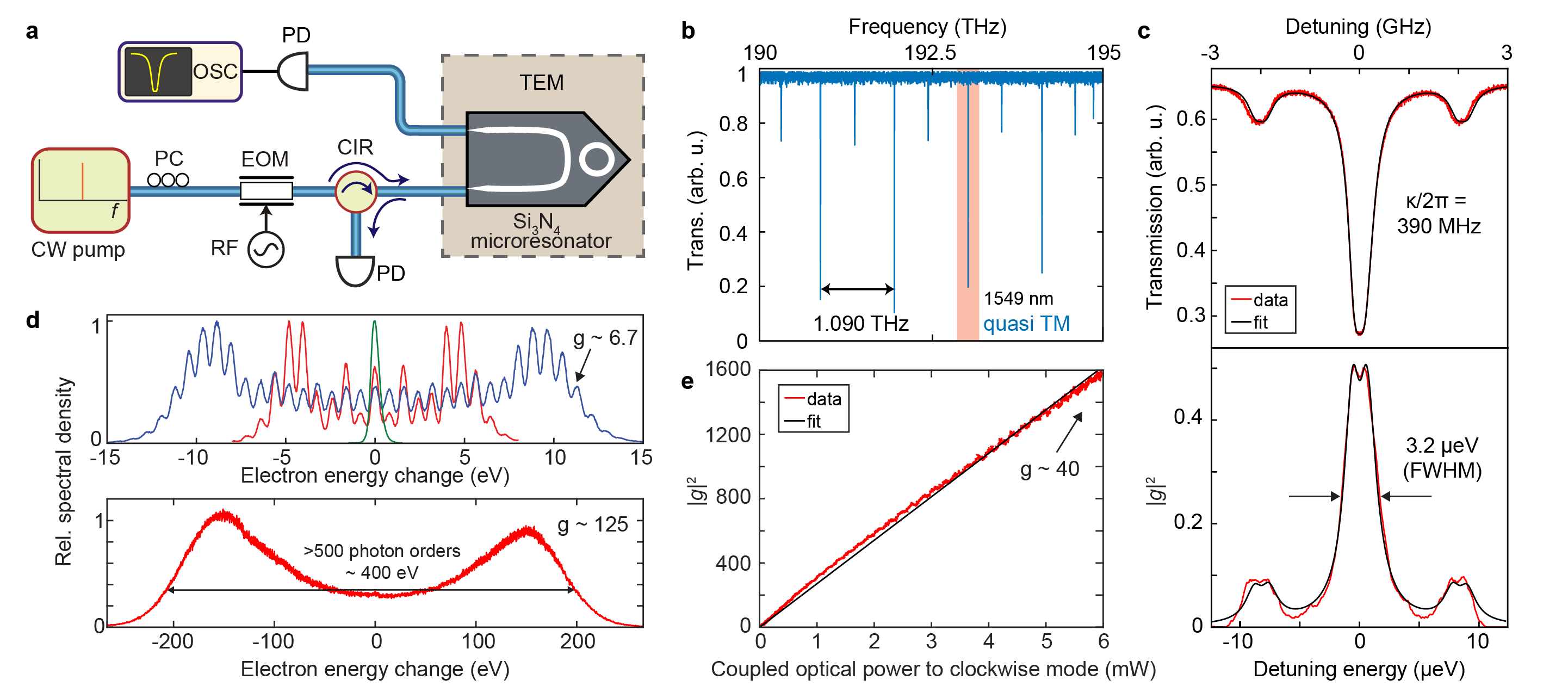}
\caption{\textbf{Simultaneous optical and electron spectroscopy of a high-Q resonator mode.} \textbf{a)} A continuous-wave (CW) external cavity laser is used to excite the quasi-TM or TE modes of the $\mathrm{Si_3N_4}$ microresonator by using a polarization controller (PC). The calibration of the relative frequency in the transmission through the photonic chip and the measured $g$ is accomplished by imparting sidebands ($\pm 2  \mathrm{GHz}$) using an electro-optic phase modulator (EOM). The back-reflected light, originating from backscattering in the microresonator, is measured using an optical circulator (CIR). (PD: photodiode, OSC: oscilloscope, TEM: transmission electron microscope, RF: radio-frequency synthesizer.) \textbf{b)} The normalized transmission scan of the microresonator quasi-TM mode measured outside of the TEM having $Q$-factor $\sim 0.77 \times 10^6$ with $ \mathrm{\kappa_0/ 2\pi~  = ~ 112~ MHz }$ and $\mathrm{\kappa_{ex}/2\pi ~ =~  139~ MHz}$. (see Methods: Optical characterization of microresonators inside TEM). The free spectral range (FSR) of the microresonator is $\sim$~1.090~THz. \textbf{c)} Simultaneously measured optical transmission at the output waveguide (top) and \textbf{$|g|^2$} retrieved from the electron energy spectra (bottom) when the electron-beam is interacting with the evanescent field of the microresonator. A slight splitting is present due to modal coupling. The measured \textbf{$|g|^2$} trace follows the power coupled to the clockwise mode. \textbf{d)} Example electron energy spectra for low (top) and high (bottom, 38~mW) optical input in the bus waveguide. \textbf{e)} The $|g|^2$ varies linearly with optical power coupled to the clockwise mode of the cavity.
} \label{fig_spectroscopy}
\end{figure*}

\section*{Integrated photonics for electron phase modulators in the continuous wave regime}
To facilitate high interaction strengths, we use photonic chip based $\mathrm{Si_3N_4}$ microresonators. $\mathrm{Si_3N_4}$ is a CMOS compatible material particularly suitable due to its high power handling capability \cite{Brasch2016}, radiation hardness \cite{Brasch2014}, and ability to achieve exceptionally low propagation losses (below 1~dB/m at 1550~nm) \cite{Pfeiffer2016,Liu2021}. Moreover they allow for dispersion engineering and efficient fiber to chip coupling \cite{Liu2018_taper}. $\mathrm{Si_3N_4}$ based integrated photonic circuits have been driving major progress in nonlinear optical devices, in particular soliton microcombs \cite{Kippenberg2018,Gaeta2019} that are already used in numerous system level applications ranging from coherent telecommunication \cite{Marin-Palomo2017} to exoplanet astrophysical spectrometer calibration \cite{Obrzud2019,Suh2019}.
The chip was fabricated using the photonic Damascene process \cite{Pfeiffer2016,Liu2021} without top oxide cladding to allow for efficient free-electron-light interaction with the evanescent field.  An optical micrograph is shown in Figure~\ref{fig_setup}d, also illustrating that the microresonator is placed close to a triangular edge of the substrate to minimize undesired electron-substrate interactions. The ring microresonator of \SI{20}{ \micro m} radius is coupled to the bus waveguide with a cross-section of 800~nm~\texttimes~650~nm to achieve an external coupling rate of $\kappa_{\mathrm{ex}}/2\pi~  =~ $\SI{130}{MHz} in order to operate close to critical coupling where the cavity power buildup is maximum.

Phase matching at different electron energies is accessible by modifying the waveguide geometry (Fig.~\ref{fig_setup}e). 
In the current study, we design a ring microresonator with dimensions \SI{2.2}{ \micro m} $\times$ 650~nm to provide phase matching at an optical frequency of $\sim$~193 THz (wavelength of $\sim$~1549~nm within the telecommunication band) for a target electron energy of 115 keV. Figure~\ref{fig_setup}f shows a FEM simulation (see Methods) of the fundamental quasi-TM mode profile of the microresonator geometry we use in our study. The field distribution of the major contributing component $\mathrm{E_{\varphi}}$, along the electron propagation direction, is shown for this design. Owing to the small mode area and considerable evanescent field component, we predict a vacuum coupling rate of $g_0/2\pi \sim\SI{e11}{Hz} $ over an interaction time of $\tau\sim \SI{3e-14}{s}$ in our system. The high-$Q$ factor (with internal $Q$-factor $Q_0 \sim 0.74 \times 10^6$, )
of the optical ring microresonator also helps us to achieve extremely low coupled optical power to reach a unity coupling constant $g\sim 1$ at $P=n_c\hbar\omega\kappa\sim\SI{1}{\micro W}$, where $\kappa/2\pi ~  = ~ \SI{390}{MHz}$ is the measured cavity decay rate inside the electron microscope and $n_c=|\alpha|^2$ is the intracavity photon number. However, we do observe a moderate degradation of the microresonator $Q$-factor (see Methods) upon insertion into the electron microscope, which could be related to some charging, strain or contamination of the resonator chip.
As is required when operating in the TEM, the photonic structure is packaged via ultrahigh numerical aperture (UHNA) fibers (see Figure~\ref{fig_setup}c) with a fiber-to-fiber coupling of $>$ 25\% (see Methods). The fibers are transferred into and out of the TEM via fiber feedthroughs, and further connected to the continuous-wave (CW) driving laser and the photodiode.

\section*{Simultaneous optical and electron spectroscopy of a photonic chip based microresonator}

\begin{figure*} [!ht]
\centering
\includegraphics{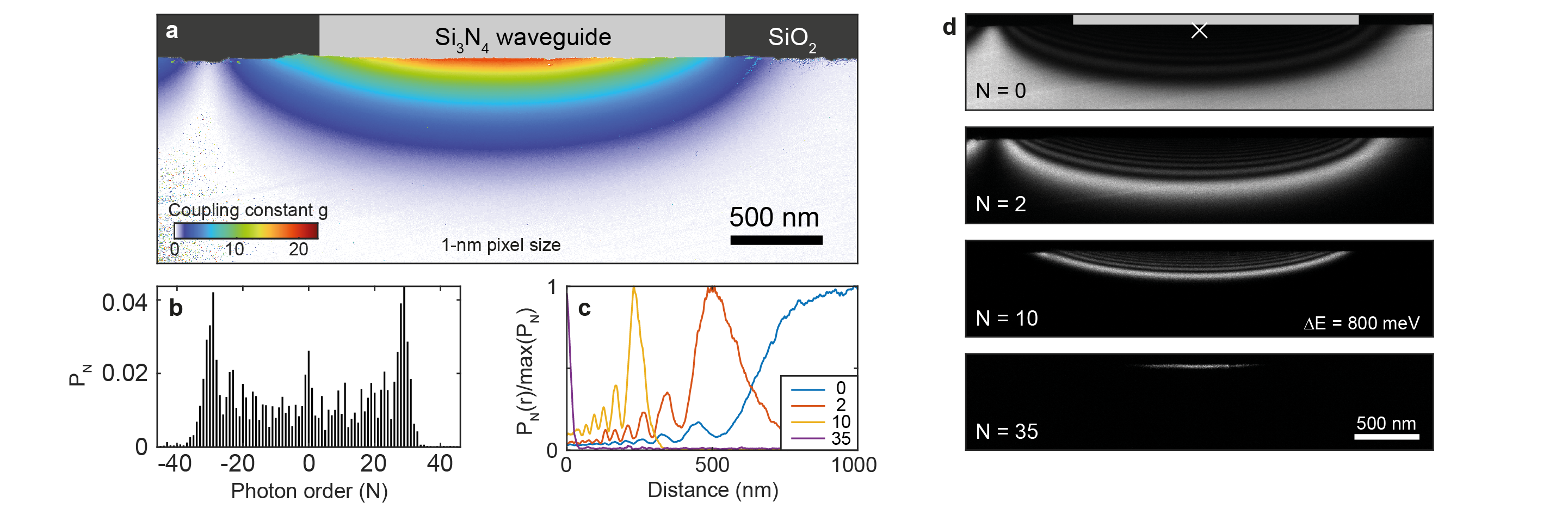}
\caption{\textbf{High-resolution hyperspectral imaging of the quasi-TM photonic-chip-based microresonator mode.} \textbf{a)} Spatial map of the coupling constant $g$ at phase-matching condition (115-keV electron energy) with a 1-nm image resolution. \textbf{b)} Exemplary electron spectrum retrieved from energy filtered spatial map (position indicated by cross in d)). \textbf{c)} Photon order dependent sideband population as a function of impact parameter (distance to the surface). \textbf{d)} Energy-filtered images of selected photon sidebands $N$ (indicated). The energy window has a width of 800~meV and the position of the microresonator is indicated in grey.}\label{fig_EFTEM}
\end{figure*}

\begin{figure*}[!ht]
\centering
\includegraphics{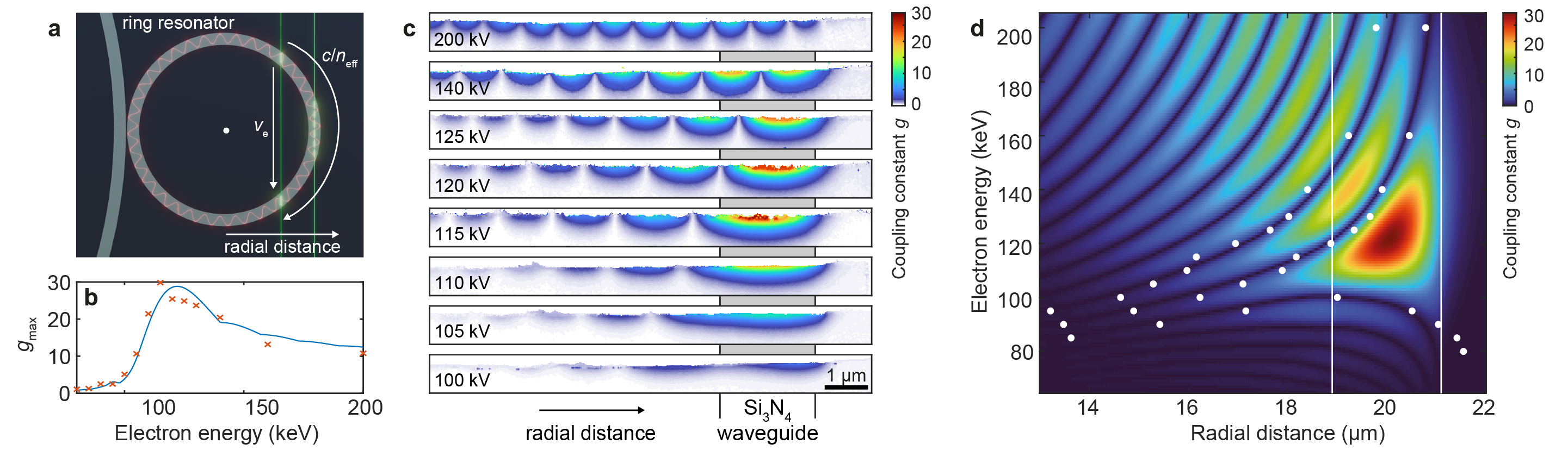}
\caption{\textbf{Phase-matching and Ramsey-type interference.} \textbf{a)} Geometry (side view) of the electron beam passing by the photonic microresonator and interacting with the quasi-TM photonic-chip-based microresonator mode
The double interaction and phase-shifts between light and electron wave lead to a characteristic spatial distribution of the coupling constant $g$ due to Ramsey-type interference. \textbf{b)} Maximum coupling parameter $g$ as a function of electron kinetic energy retrieved from experiment (red dots, c)) and numerical simulations (blue line, d)). \textbf{c)} Quantitative $g$-maps retrieved for variable electron energy, showing the varying spatial distribution and amplitude of the electron-light interaction. The radial position of the $\mathrm{Si_3N_4}$ waveguide is indicated. \textbf{d)} Numerical simulation of $g$ as a function of lateral position (just above the surface) and electron energy. The white dots overlay the experimentally observed minima as retrieved from c) (80-200~kV electron beam energy). The white lines mark the waveguide position.}\label{fig_HT}
\end{figure*}

In the experiments, the fiber-coupled microresonator is driven by a 1550~nm continuous-wave laser via the bus waveguide. It is mounted on a custom-designed sample holder and placed in the object plane of the field-emission TEM (Fig.~\ref{fig_setup}a, see Methods). Parallel to the surface, the electron beam passes by the waveguide and interacts with the confined optical mode (Fig.~\ref{fig_setup}a inset). After traversing the structure, the electron kinetic energy distribution is characterized with an imaging electron spectrometer in two different ways. Specifically, high-dispersion electron energy spectra are recorded by positioning a sharply focused electron beam in front of the microresonator. Alternatively, energy-filtered transmission electron microscopy (EFTEM) using collimated TEM-illumination is used for imaging the interaction across the entire cavity mode near-field. While the former yields electron spectra for varying experimental parameters (Fig.~\ref{fig_spectroscopy}), the latter enables imaging of individual sideband populations with high spatial resolution (see Figs.~\ref{fig_EFTEM}\&\ref{fig_HT}).

We first investigate the strength of the coupling parameter $g$ using a focused electron probe, recording electron spectra while scanning the laser over a photonic microresonator resonance employing the setup depicted in figure \ref{fig_spectroscopy}a. The transmission spectrum, displayed in figure \ref{fig_spectroscopy}, shows the quasi-TM microresonator modes spaced by a free spectral range (FSR) of 1.090~THz.
In order to calibrate the optical frequency, an electro-optical modulator (EOM, driven at 2.0~GHz) is used to generate sidebands that can be observed in the transmission spectrum.
Figure~\ref{fig_spectroscopy}c shows simultaneous optical and in-situ electron spectroscopy of the laser-excited microresonator mode. The laser is tuned to a single optical mode at $\sim$1549.4~nm ($\kappa / 2 \pi$~$~ =~ $~390 MHz, see Methods) and the focused electron beam (120-keV beam energy, 25-nm focal spot size, 1-mrad convergence semi-angle) is centered just above the surface of the microresonator (see Fig.~\ref{fig_EFTEM}) to record electron spectra for a stationary beam. Harnessing the high-$Q$ intracavity enhancement, we observe strong populations $P_N$ in multiple photon orders $N$, reaching a previously inaccessible regime for a continuous laser light source and electron beam (Fig.~\ref{fig_spectroscopy}d). The coupling parameter $g$ is retrieved from the spectra (see Methods), and we find the expected linear increase of $|g|^2$ with the optical power coupled to the clockwise-propagating mode, determined from the recorded optical transmission and reflection data (Fig.~\ref{fig_spectroscopy}e).

We achieve a coupling, with $g \sim 1.2$ and $\sim 40$ for about \SI{6}{\micro W} and 6~mW, respectively, of optical power $P$ coupled into the interacting microresonator mode.
Increasing the power in the bus waveguide to $\sim \unit[38]{mW}$, the interaction strength can be increased to $g \sim 125$ corresponding to the generation of $> 500$ photon sidebands (see Figure \ref{fig_spectroscopy}d).
The spectral line shape of the resonance is analyzed by simultaneously recording the transmitted optical power leaving the chip and extracting $g$ for varying frequency detuning (Fig.~\ref{fig_spectroscopy}c).
The optical transmission trace, containing power coupled from the clockwise mode interfering with the input light, shows a trace with a full-width-half-maximum of 560~MHz (total line width $\kappa/2\pi$~ =~ 390~MHz, see SI).
Interestingly, the electron spectra, sensitive only to the intra-cavity power stored in the clockwise-rotating optical mode, display a double-peaked structure originating from coupling optical power to the frequency degenerate counter-clockwise optical mode \cite{Kippenberg2002} (see SI, Figure S2). These data demonstrate continuous-wave electron energy gain spectroscopy (EEGS) \cite{Asenjo-Garcia2013} at an extraordinarily narrow spectral feature of only \SI{3.2}{-\micro eV}  in width. The concept can be transferred to arbitrary high-$Q$ optical modes, with previously demonstrated spectral linewidths below $\kappa/2\pi\sim$ 10~MHz \cite{Liu2021}.

\section*{Mode imaging and phase matching}

The implementation of photonic chip based high-$Q$ optical microresonators as efficient electron phase modulators requires insights into the real-space scattering distribution. Therefore, we conduct high-resolution, energy-filtered spatial mapping with a collimated electron beam (Fig.~\ref{fig_EFTEM}a). Employing an imaging energy filter with a sharp bi-sided window of 800~meV width, equivalent to the energy of 1549~nm photons, spatial maps in the object plane are recorded for the discrete sideband populations $P_N$ (Fig.~\ref{fig_EFTEM}d, see Methods). This way, at each image pixel a photon-order resolved spectrum is extracted (Fig.~\ref{fig_EFTEM}b) and the coupling parameter $g$ is retrieved. Previously encountered signal-to-noise ratio limitations in PINEM are lifted by the available high current in continuous electron beams (about 3-4 orders of magnitude higher electron flux), enabling a high-fidelity quantitative reconstruction of the local coupling with a 1-nm spatial resolution (Fig.~\ref{fig_EFTEM}a). To account for the large number of spectra comprising an image, systematic data clustering is applied to retrieve large $g$-maps (see Methods). The distance dependent sideband occupations (Fig.~\ref{fig_EFTEM}c) reveal a strong and high-contrast modulation, indicating the absence of spatial and temporal averaging often encountered in PINEM.

The microresonator waveguide is designed to match the phase velocity of the excited optical mode with the group velocity of electrons at an energy around $\sim$115~keV. Figure~\ref{fig_HT} shows the spatial pattern of the electron-light coupling near the ring-shaped resonator (Fig.~\ref{fig_HT}a). As described above, quantitative $g$-maps are extracted by energy-filtered imaging for a varying acceleration voltage of the electron microscope (Fig.~\ref{fig_HT}c), revealing an oscillatory modulation of $g$ along the chip surface, as well as an amplitude change with electron energy (Fig.~\ref{fig_HT}b). The interaction region near the outer radius of the ring resonator allows for efficient and spatially extended phase modulation of the electron beam (tangent line trajectory) with an amplitude governed by the electron-energy-dependent phase mismatch. Around the optimum coupling, the spatial distribution of $g$ closely resembles the electric field configuration of the excited quasi-TM mode at 1549~nm (see Fig.~\ref{fig_EFTEM}a). Beam trajectories further to the center of the ring resonator (secant line) are governed by two sequential interactions with the microresonator mode. This results in Ramsey-type constructive or destructive interference and strong spatial modulations of $g$ that depend on the relative phase of both individual interactions \cite{Echternkamp2016}.
Numerical simulations (Fig.~\ref{fig_HT}d, see Methods) based on the FEM simulation of the optical mode profile and taking into account the electron trajectory through the near-field as well as the resonator power enhancement, closely reproduce the experimental observations and show a good match in the position of the minima in $g$ (white dots in overlay). Furthermore, excellent agreement is achieved for dependence of the electron kinetic energy on the maximal observed value of $g$ (Fig.~\ref{fig_HT}b). A steep cutoff in electron-light coupling is found below an electron energy of $\sim$100~keV, because, in contrast to nanoscale-confined optical fields, the optical mode is devoid of higher-momentum electric field components. Besides enhancing the total coupling, an increase in microresonator radius will further sharpen the phase matching condition in energy.

\section*{Conclusion}
In summary, we demonstrate highly efficient, phase-matched and single-optical-mode interaction of free electrons with photonic chip based high-$Q$ microresonators, driving transitions at micro-Watt continuous optical pump power. Based on simultaneous in-situ optical and electron measurement, the cavity QED type setting yields a quantitative understanding of the interaction and electron energy gain spectroscopy at the \SI{}{ \micro eV} level.  
This approach can facilitate a seamless integration of temporal phase plates and attosecond metrology in future electron microscopy applications, promoting locally driven coherent radiation and excitations, thus defining new directions in coherent electron spectroscopy. We envisage various future architectures featuring coupled resonators for advanced electron state preparation and readout. Furthermore, we believe that the unprecedented degree of control represents a key step towards free-electron cavity quantum electrodynamics. This promises electron-heralded single-photon sources or electron imaging enhanced by photon post-selection, and facilitates cavity-mediated electron-photon and electron-electron entanglement.

\emph{Note: A study presenting continuous-wave IELS in the context of dielectric laser acceleration was posted at same time as our work. \cite{Dahan2021}}

  \close@column@grid
  \cleardoublepage
  \twocolumngrid

\part*{Methods}
\medskip
\begin{footnotesize}
\section{Numerical simulations}
The effective refractive index ($n_\mathrm{eff}$) and the mode profile of the $\mathrm{Si_3N_4}$ microresonator are calculated via finite-element-method simulations (COMSOL Multiphysics) (see Fig.~\ref{fig_setup}f). Mode analysis is performed for the two-dimensional axially symmetric model with the cross-section of the microresonator. The $\mathrm{Si_3N_4}$ core has a rectangular cross-section with width W and height H. The microresonator has a $\mathrm{SiO_2}$ bottom cladding and an air (vacuum) top cladding. The refractive index of $\mathrm{Si_3N_4}$ used in the simulation is obtained from in-house ellipsometry measurement (J.A. Woollam), and the refractive index of $\mathrm{SiO_2}$ is obtained from ref.~\cite{Malitson1965}.
From the electric field distribution in the cross-section ($r,~z$), the complex electric field along the electron trajectory $E_{\text{traj}} (r,z,y)$, related to the physical field by $E_\mathrm{physical}=\mathrm{real}[E_{\text{traj}}]$, can be determined via $E_{\text{traj}} (r,z,y) = E_{\varphi}(r,z) \cos(y/r) + E_r \sin(y/r)$. The coupling constant $g(r,z,E)$ is then found by evaluating $g = \frac{e}{2 \hbar \omega} \int_{-\infty}^{\infty} E_{\text{traj}}(r,z,y) \exp(-i \frac{\omega}{v_e} y) \text{d} y$ \cite{Park2010,Feist2015} for different electron energies $E$, where $v_e = c \sqrt{1- \nicefrac{1}{\left(1+\nicefrac{E}{m_0 c^2}\right)^2}}$ is the relativistic electron velocity.

\section{fiber-integrated silicon nitride microresonators}
The designed microresonators are fabricated using the photonic Damascene process \cite{Pfeiffer2016,Liu2021}. 
A 4-inch silicon substrate with  \SI{4}{ \micro m}  thick thermal wet SiO$_2$ cladding is used. 
The substrate is coated with deep-ultraviolet (DUV) photoresist, and microresonators and bus waveguides are patterned on the substrate via DUV stepper photolithography (248~nm KrF excimer laser). 
The pattern is then dry-etched into the SiO$_2$ cladding using C$_4$H$_8$/O$_2$/He etchants, to create waveguide preforms.
Stoichiometric Si$_3$N$_4$ film is deposited on the patterned substrate via low-pressure chemical vapor deposition (LPCVD), to fill the waveguide preforms and to form the waveguide cores.
Afterwards, etchback and chemical-mechanical polishing \cite{Liu2021} are used to planarize the substrate and to remove excess Si$_3$N$_4$.
The entire substrate is further annealed at 1200$^\circ$C to remove the residual hydrogen content in Si$_3$N$_4$, to reduce hydrogen-induced absorption loss in Si$_3$N$_4$ waveguides. 
This high-temperature annealing is critical to achieve high microresonator $Q$ at telecommunication bands around 1550~nm. 
No top SiO$_2$ cladding is added on the Si$_3$N$_4$ waveguides, such that the electron beam can interact with the optical mode.
Finally, photolithography with alignment is used to precisely position the Si$_3$N$_4$ microresonator close to the chip edge (within a distance smaller than  \SI{10}{\micro m}, key for aligning the electron beam to the Si$_3$N$_4$ microresonator).
Deep reactive-ion etching is used to separate the entire substrate into hundreds of individual dies / chips for the following die-level integration with fibers.

The light is coupled into and out of the $\mathrm{Si_3N_4}$ photonic chip on one edge, while the microresonator is placed on the other edge. To avoid cutting the electron beam by the substrate (see Fig.~\ref{fig_setup}c), the near edge is clipped and forms a tight supporting triangle. A  bus waveguide (800~nm~\texttimes~650~nm) is used to evanescently couple the light into the multimode microresonator (\SI{2.2}{ \micro m} $\times$ 650~nm ) to achieve a better coupling ideality \cite{Pfeiffer2017}. The inverse-taper waveguides \cite{Liu2018_taper}  are used to facilitate efficient light coupling via ultra-high numerical aperture (UHNA-7) fibers (mode field diameter $\sim$ \SI{3.2}{ \micro m} at 1550~nm). A 2-3~cm long UHNA-7 fiber having a thermally expandable core is spliced to standard single-mode fiber (SMF-28) with a splicing loss of $<$~0.2~dB. A chip through coupling efficiency (fiber-chip-fiber) of $>$~25~\% is achieved using the UHNA-7 fiber. The photonic packaging is done by first aligning the UHNA-7 fiber via a custom-built holder to optimize the coupling. Then, a small drop of epoxy ($\sim$ \SI{150}{ \micro m}, Fig.~\ref{fig_setup}b) is dispensed using a precise pneumatic valve and cured using a UV lamp in four small time steps ($\sim$ 2 to 5 minutes). A long-term ($\sim$ 1-2 days) coupling stability is performed at low optical power by monitoring the transmitted light. The broadband characterization of the microresonators is done by employing a widely tunable diode laser. The transmission spectrum is calibrated using a self-referenced optical frequency comb and Mach-Zehnder interferometer (MZI) \cite{Liu2016}. The resonance is fitted using models explained in the SI (fitting model) to extract $\kappa_0 / 2 \pi$ (intrinsic loss rate), $\kappa_{\mathrm ex} / 2 \pi$ (external coupling loss rate), and  $\gamma / 2 \pi$ (mode  coupling rate) for both quasi-TE and quasi-TM mode families. The mean intrinsic linewidth is $\sim$ 110-120 MHz ($Q_0 \sim$ $1.75\times 10^6$) for quasi-TM mode family. The cavity resonance center at $\sim$ 1549~nm is critical coupled with $ \mathrm{\kappa_0/ 2\pi ~ =~  112~ MHz }$ and $\mathrm{\kappa_{ex}/2\pi ~ =~  139~ MHz}$ ($Q\sim 0.77 \times 10^6$). 

 \section{Optical characterization of microresonators inside the TEM}
The packaged sample is transferred into the TEM using a custom holder with vacuum fiber feedthroughs. The fibers connected to the microresonator chip are fed through the hollowed pipe part of the holder and the T-shaped base holding the photonic chip is mounted on an adapter that allows for a placement of the entire structure in the sample region of the TEM. 
The optical setup used to perform the spectroscopy in the TEM is shown in Fig.~\ref{fig_spectroscopy}a. The setup is driven by a CW-laser (Toptica CTL 1550), with a maximal power of 40~mW, a tunable wavelength from 1510 nm to 1630 nm, and a linewidth below 10 kHz. The laser is coupled to an electro-optic modulator (EOM) that allows for frequency calibration of the transmission scan and the dependence of $g$ on the detuning from the resonance frequency.  A polarization  controller is used to align the input light to either the quasi- TE or TM mode of the microresonator. An optical circulator is used to probe the light reflected from the microresonator due to bulk and Rayleigh surface scattering responsible for the splitting of clockwise and counterclockwise  cavity modes. The optical transmission and reflection  are measured simultaneously to calibrate intracavity power dissipated in the clockwise mode (see SI). We note that a degradation of the Q-factor from $0.77 \times 10^6$ to $0.49 \times 10^6$, resulting in an additional increase in total linewidth ($\sim$ 150 MHz), is observed when the sample is transferred into the TEM, that might be due to hydrocarbon deposition. Theory predicts that $|g|^2$ is proportional to the photon number  $n_\mathrm{cw}$ of the clockwise mode optical cavity mode. The intracavity photon numbers are varied by sweeping the laser frequency through the optical resonance to find the $|g|^2$ dependence on optical power. The $g(t)$ value is extracted for  each time step of the sweep, $t$, by fitting the measured electron spectrum, and the intracavity photon number $n_\mathrm{cw}(t)$ is measured from the optical transmission signal using the steady state Langevin equation. By plotting $|g(t)|^2$ against coupled optical power in the clockwise mode $P=\kappa\hbar\omega n_\mathrm{cw}(t)$, we obtain the power sweep curve shown in Fig.~\ref{fig_spectroscopy}(e). More detailed data analysis procedure and the analytical expression of various field quantities can be found in the SI.

\section{Electron microscope and experimental setup}
The in-situ experiments are performed in the Göttingen UTEM \cite{Feist2017}, which is based on a thermal field-emission TEM (JEM 2100F, JEOL Ltd.). The electron gun is operated in the extended Schottky regime, yielding a continuous electron beam with an initial energy spread of 0.5~eV at a variable electron energy from 80 to 200~keV and typical beam currents of 10-50~pA in the sample region. Preventing clipping of the electron beam at the extended microresonator chip ($\sim$\SI{30}{ \micro m} sample height at the probing position), a low beam convergence is set by turning off the magnetic objective lens in low magnification (LM) TEM and STEM mode, for high-resolution imaging and spectroscopy, respectively. For analyzing the electron beam passing the microresonator structure, a post-column imaging energy filter (CEFID, CEOS GmbH) is employed, equipped with a scintillator-coupled CMOS camera (TemCam-XF416ES, TVIPS GmbH; used for all measurements, if not mentioned otherwise) and a hybrid-pixel electron detector (EM CheeTah T3, Amsterdam Scientific Instruments B.V.). The sample tilt is adjusted carefully to align the electron beam parallel to the surface of the resonator chip, probing the local electron-light interaction at the microresonator.

\section{Local spectroscopy of the resonator mode}
Electron spectroscopy (Fig.~\ref{fig_spectroscopy}) is implemented in LowMAG STEM mode (indicated magnification: 1000x) at 120~keV electron energy, achieving an electron focal spot size of about 25~nm with a beam semi-convergence angle of 1.1~mrad (\SI{100}{ \micro m} condenser aperture, Fig.~\ref{fig_spectroscopy}d) and 0.45~mrad ($\sim$\SI{40}{ \micro m} condenser aperture, Fig.~\ref{fig_spectroscopy}e,f), respectively. The STEM focal plane is set to the middle of the ring-resonator and the lateral beam position relative to the chip is software controlled (Filter Control, CEOS GmbH) via an external scan generator (USG, TVIPS GmbH). Single spectra are recorded with an integration time of 100~ms (Fig.~\ref{fig_spectroscopy}d). Electron spectroscopy with fast sweeps of the laser frequency is captured by an event-based hybrid pixel detector (EM CheeTah T3, Timepix3 ASIC, 1.56-ns timing precision) and discrete binning in time (\SI{100}{\micro s} windows, Fig.~\ref{fig_spectroscopy}c,e). For each spectrum, the electron-light coupling constant $g$ is extracted by fitting the shape and amplitudes of the individual spectral sidebands $N$ with a comb of Voigt-peaks and normalized occupations $P_N=J_N(2|g|)^2$. For details on the fitting procedure, see \cite{Feist2015, Kfir2020}.

\section{Energy-filtered imaging and quantitative {\it \lowercase{g}}-maps}
Quantitative maps of the electron-light coupling constant $g$ are retrieved by energy-filtered transmission electron microscopy (EFTEM) imaging (Figs.~\ref{fig_EFTEM} and \ref{fig_HT}), resolving individual photon sidebands with a double-sided energy window of only 800~meV width and typical bi-sided r.m.s. non-isochromaticity (energy window edge sharpness) of about 30~meV (8~mm field-of-view of the spectrometer entrance aperture plane, for technical details, see \cite{Kahl2019}). By synchronized change of the microscope's high tension (step size of 0.8~V) relative to the center beam energy (80-200~keV), images $I_N(x,y)$ are recorded that contain inelastically scattered electrons, losing or gaining an integer number $N$ of photon energies $\hbar\omega$=0.8~eV (Fig.~\ref{fig_EFTEM}c). For each spectral sideband an image is recorded with 1-s integration time, resulting in EFTEM datasets containing up to 121 image slices, enabling a quantitative retrieval of the coupling constant $g$ at each image pixel (see Fig ~\ref{fig_EFTEM}d)). Accounting for a large number of image positions (up to 4096x4096 pixels in Fig.~\ref{fig_EFTEM})), the data is collocated into 1024 subsets using the $k$-means++ clusters algorithm (implemented in MATLAB 2020b, MathWorks Inc.), with the sideband occupations as the input vector. The resulting centroid locations (metric: squared Euclidean distances) give a set of averaged spectra. For each of these, the coupling constant $g$, relative uncertainty $dg/g$ (with standard deviation $dg$ of Gaussian distribution in $g$) and spectral amplitudes $A$ are fitted to the recorded image stacks $I_N(x,y)$. The averaged intensity spread of the ZLP to neighboring energy windows, $B$, is accounted for by applying a convolution $I_N(x,y) = A(x,y) \cdot P_N(x,y) \ast B$ (at 115-keV beam energy, the 800-meV energy window contains about 82\% of the ZLP intensity). Furthermore, remaining ZLP intensity at stronger coupling ($g>3$), resulting from limited rejection contrast of the spectrometer slits and scattering at the entrance aperture, is disregarded (about 2(0.8)\% diffuse background for $N=0(\pm1)$). Finally, the retrieved fitting parameters are assigned to the individual pixels before clustering, resulting in a quantitative map of the electron light coupling constant with 1-nm spatial resolution (Figs.~\ref{fig_EFTEM} and \ref{fig_HT}).

\section*{Data Availability Statement} 
The code and data used to produce the plots within this work will be released on the repository \texttt{Zenodo} upon publication of this preprint.

\section*{acknowledgments} 
We thank Hugo Louren\c{c}o-Martins, Itay Shomroni, Johann Riemensberger and Nils Johan Engelsen for useful discussions. We thank Jijun He for taking photographs of the photonic chips, John Gaida for support with the TEM holder design and Murat Sivis for making the render graphics.

\noindent\textbf{Funding Information:} All samples were fabricated in the Center of MicroNanoTechnology (CMi) at EPFL. This material is based upon work supported by the Air Force Office of Scientific Research under award number FA9550-15-1-0250.  This work was further supported by the Swiss National Science Foundation under grant agreements 185870 (Ambizione), 182103 and 176563 (BRIDGE). 
The work at the Göttingen UTEM Lab was funded by the Deutsche Forschungsgemeinschaft (DFG, German Research Foundation) through  432680300/SFB\,1456 (project C01) and the Gottfried Wilhelm Leibniz program, and the European Union’s Horizon 2020 research and innovation programme under grant agreement No. 101017720 (FET-Proactive EBEAM). Y.Y. acknowledges support from the EU H2020 research and innovation program under the Marie Sklodowska-Curie IF grant agreement No. 101033593 (SEPhIM). O.K. acknowledges the Max Planck Society for funding from the Manfred Eigen Fellowship for postdoctoral fellows from abroad.

\noindent\textbf{Author contribution:}
A.S.R. and J.L. designed the photonic chip devices. 
J.L. and R.N.W. developed the fabrication process and fabricated the devices. 
A.S.R. characterized and packaged the devices. 
J.W.H., G.H., J.P. and Y.Y. performed numerical simulations.
G.H., O.K. and C.R. devised the theory section.
J.W.H. and A.F. carried out the the TEM experiments and data analysis, assisted by G.A., F.J.K. and M.M..
The study was planned and directed by O.K., C.R. and T.J.K.
The manuscript was written by J.W.H., A.S.R., A.F., G.H., Y.Y., C.R., T.J.K., after discussions with and input from all authors.

\end{footnotesize}
\bibliographystyle{apsrev4-2}

\bibliography{manuscript_arXiv}

\end{document}